%% file: paper.tex
\documentclass{sig-alternate-05-2015}

\input{custom.tex}

\begin{document}

\CopyrightYear{2016}
\setcopyright{acmcopyright}
\conferenceinfo{ACSAC '16,}{December 05-09, 2016, Los Angeles, CA, USA}
\isbn{978-1-4503-4771-6/16/12}\acmPrice{\$15.00}
\doi{http://dx.doi.org/10.1145/2991079.2991103}

\title{EvilCoder: Automated Bug Insertion}

\subtitle{}

\numberofauthors{2}

\author{
\alignauthor
Jannik Pewny\\
       \affaddr{Horst G\"ortz Institut (HGI)}\\
       \affaddr{Ruhr-Universit\"at}\\
       \affaddr{Bochum, Germany}\\
       \email{jannik.pewny@rub.de}
\alignauthor
Thorsten Holz\\
       \affaddr{Horst G\"ortz Institut (HGI)}\\
       \affaddr{Ruhr-Universit\"at}\\
       \affaddr{Bochum, Germany}\\
       \email{thorsten.holz@rub.de}
}
\date{June 1st, 2016}

\maketitle

\begin{abstract}
\input{sections/abstract.tex}
\end{abstract}

\input{sections/introduction.tex}

\input{sections/approach.tex}
\input{sections/implementation.tex}

\input{sections/evaluation.tex}

\input{sections/discussion.tex}

\input{sections/related_work.tex}

\input{sections/conclusions.tex}

\input{sections/acknowledgement.tex}

\bibliographystyle{abbrv}
\bibliography{sigproc}  %

\end{document}

%% file: custom.tex
\usepackage[usenames,dvipsnames]{color}
\usepackage{graphicx}
\usepackage{enumerate}
\usepackage{amssymb}
\usepackage{url}
\usepackage{tabularx}
\usepackage{amsmath}
\usepackage{hyperref}
\usepackage{listings}

\usepackage{caption}
\DeclareCaptionType{copyrightbox}
\usepackage{subcaption}
\usepackage{colortbl}
\usepackage{multicol}
\usepackage{xcolor}
\usepackage{dblfloatfix}
\usepackage{cite}
\usepackage{paralist}
\usepackage{csquotes}

\usepackage{multirow}

\newcommand{\tool}{\textsc{EvilCoder}}

%% file: sections/abstract.tex
The art of \emph{finding} software vulnerabilities has been covered extensively
in the literature and there is a huge body of work on this topic. In contrast, the
intentional \emph{insertion} of exploitable, security-critical bugs %
has received little (public) attention yet.
Wanting more bugs seems to be
counterproductive at first sight, but the comprehensive evaluation of bug-finding
techniques suffers from a lack of ground truth and the scarcity of bugs.

In this paper, we propose \tool{}, a system to automatically find potentially
vulnerable source code locations and modify the source code to be actually
vulnerable. More specifically, we leverage automated program analysis
techniques to find sensitive sinks which match typical bug patterns (e.g., a
sensitive API function with a preceding sanity check), and try to find
data-flow connections to user-controlled sources. We then transform the source
code such that exploitation becomes possible, for example by removing or
modifying input sanitization or other types of security checks.  Our tool is
designed to randomly pick vulnerable locations and possible modifications, such
that it can generate numerous different vulnerabilities on the same software
corpus.
We evaluated our tool on
several open-source projects such as for example \texttt{libpng} and \texttt{vsftpd},
where we found between 22 and 158 unique connected source-sink pairs per project.
This translates to hundreds of potentially vulnerable data-flow paths
and hundreds of bugs we can insert.
We hope to
support future bug-finding techniques by supplying freshly generated,
bug-ridden test corpora so that such techniques can (finally) be evaluated and
compared in a comprehensive and statistically meaningful way.

%% file: sections/introduction.tex
\section{Introduction}
\label{introduction}

Many different kinds of software vulnerabilities exist, ranging from simple
buffer overflows~\cite{stack-smashing} over integer overflows~\cite{integer-overflow} to temporal
errors~\cite{temporal-bugs} or even errors introduced by the compiler due to
undefined behavior~\cite{undefined-behaviour}. Numerous approaches exist for \emph{finding}
such vulnerabilities and there is a huge body of work on this topic. Such
techniques are based on an analysis of the source code (e.g.,~\cite{Livshits2005, redebug, yamaguchi:2014}) or based on
binary analysis (e.g.,~\cite{binary1, binary2}), while others leverage fuzz testing (e.g.,~\cite{BlackBoxFuzzing, fuzzing1, fuzzing2}) or other techniques (e.g.,~\cite{symbolic-exec, cross-arch}).

We think that evaluating different approaches for finding
vulnerabilities is a hard problem in practice for two
main reasons: first, security vulnerabilities are scarce. That is not to say
that there are not too many of them, but for a statistically meaningful
evaluation, they are simply spread too thin and do not expose all facets of
the underlying problem. The second reason stems from the current practice to
regard finding new vulnerabilities as the most convincing argument 
for a newly proposed technique.  While this is
the sole reason as to \emph{why} we develop such
techniques, we argue that it should not be how we \emph{evaluate} bug-finding
techniques. %

In this work, we thus focus on the opposite problem: instead of finding or
fixing vulnerabilities, we study the \emph{insertion} of security-critical bugs
in complex software systems.
Compared to
the volume of source code or binary code, vulnerabilities are relatively rare
and tend to be fixed once they are found, which makes statistical evaluation
complicated at best. Having public bug-ridden test corpora with known bug
locations would help immensely to evaluate different techniques in an objective
manner. For example, the field of machine-learning proceeds like this for
years, having a few standard corpora (like the Texas Instruments-MIT speech
corpus~\cite{timit} or the Wall Street Journal corpus~\cite{wsj}), which are widely
used to compare new methods.  We are convinced that having \emph{freshly
generated} test corpora is crucial for bug-finding techniques, as it prohibits
memorizing the bug database and forces any approach to abstract from the
details of the individual bugs. This may be one of the reasons why the
available \textit{static} test corpora (e.g.,~\cite{samate, webgoat, securibench})
are so seldom used in evaluating new techniques.
Note that by generating the bugs,
one achieves \emph{ground truth} for the test corpus,
as the full vulnerable path is known.
Furthermore, such inserted bugs form a \emph{lower bound} for the number of bugs every
bug-finding approach has to find.
This can be understood in two ways: First, the approach could obviously
find more bugs in the test corpus, as the program may have had vulnerabilities
to begin with.
Second, as these ground-truth bugs are generated in an automatic way,
they are especially important to find, given that they could theoretically
be inserted by any attacker without much effort.
Having the ability to check for places where a certain bug-class might
instantiate could also be used to assess two very important data points: First,
the number of places where it could occur in the wild, which immediately gives
a hint on the exoticness of that class.  Second, an estimation of the ratio of
how often it \emph{could} occur and how often it \emph{does} occur.  This in
turn could help to prioritize aid for developers to prevent those mistakes that
are made very often.

Our approach of automatically inserting exploitable software vulnerabilities in
the source code of a given application works as follows. In a first step, we analyze the source code
in order
to find sensitive sinks, which match our supported bug patterns, and try to
find data-flow connections stemming from user-controlled sources. Then, we trace the
control flow between each source and sink connected by the data flow, which effectively points us to locations
which might hinder exploitation (e.\,g., input sanitization routines or
security checks). If our analysis indicates that we indeed found a potential
location for a vulnerability, we transform the source code such that
exploitation becomes possible. Usually, there is more than one such location
and the transformation can happen in more than one way. Hence, we can randomly
choose a variant, which leads to a large set of possible bugs.
Note that we will often use the word \emph{bug} when referring to a \emph{vulnerability},
although, one vulnerability may necessitate the insertion of multiple bugs and,
in general, not every bug is security critical. %
Nevertheless, the bugs we introduce are meant to be exploitable.
However, as the
automated creation of proof-of-concept exploits is a very complicated matter on
its own, we take care to make the bugs security critical and potentially
exploitable by design, without checking the general
satisfiability of exploitation conditions.

We implemented a tool called \tool{} that demonstrates the practical feasibility of our bug insertion techniques.
To this end, we extended \textsc{Joern}, an open-source platform for robust C/C++ analysis
by Yamaguchi et al.~\cite{yamaguchi:2014}, to facilitate interprocedural analysis and
our bug insertion techniques. In the current prototype, we focus on the
generation of taint-style vulnerabilities~\cite{yamaguchi:2014}, which cover many
vulnerability categories such as buffer overflows, integer overflows,
information leaks, and format string vulnerabilities. However, we deem
generating bug patterns for temporal vulnerabilities such as race condition or
use-after free vulnerabilities to be possible as well. 

We evaluate our tool by
applying it to four different open-source projects, where we found between
22 and 158 unique connected source-sink pairs per project.
Since each such pair can be connected with multiple data flows, each such pair
usually accounts for dozens of potential locations to insert vulnerabilities.
Including our varying source-code modifications, this could lead to hundreds of
test cases for a bug-finding tool.  To justify our claim of generating
potentially exploitable bugs, we re-generated the vulnerability of a
non-trivial exploitable CVEs from the patched version of the program.
We plan to publish our tool
and artificially bug-infested corpora
to encourage both the creation of public benchmarks for future bug-finding research
and new models for automated bug insertion.

\smallskip \noindent
In summary, our main contributions in this paper are: %
\begin{compactitem}
  \item We present a method to automatically insert security-critical bugs into complex software systems.
  \item We implemented a prototype of our techniques called \tool{}, which can insert taint-style vulnerabilities using six different classes of instrumentation.
  \item We empirically demonstrated the capabilities of our tool by finding between 22 and 158 unique, connected source-sink pairs in four open-source projects, which translates to hundreds of potential bugs.
Furthermore, we show that we can automatically re-generate the vulnerability in a non-trivial exploitable CVE from the patched version of the program.
\end{compactitem}

%% file: sections/approach.tex
\section{Approach}
\label{approach}

In this section, we define our goals and explain the design and workflow of our
approach to automatically add security-critical bugs to arbitrary
applications.

\subsection{Purpose and Scope}

Ultimately, we want to insert security-critical bugs into an application.
As noted above, we focus on generating test corpora for bug-finding techniques.
Since we do not want to exclude techniques which rely on
source code, we chose to work on %
the source code level.
Note that this does not limit binary-based approaches, as the instrumented source
code can be compiled to generate a binary executables for different processor architectures.

This choice entails the need to choose a programming language which we want to
support for a prototype implementation. Because of its widespread use in
important software and its affinity for security-critical bugs, we opted for
the C programming language.
However, we think that the general idea is applicable to other %
programming languages as well.

\subsection{Supported Vulnerability Classes}

Naturally, both finding potentially vulnerable source code locations and
instrumenting them to be actually security-critical, are specific to the class
of the vulnerability we want to cover.
We opted to focus on \emph{taint-style
vulnerabilities}~\cite{yamaguchi:2014} because they account for many different types of
vulnerabilities, mainly from the spatial domain. These vulnerabilities are
essentially characterized by an improperly secured data flow from a
user-controlled source to a sensitive sink.

While we think that supporting taint-style vulnerabilities is sufficient for a
prototype,
our system can be extended to support other kinds of vulnerabilities in future work.
Especially temporal
vulnerabilities such as security-critical race conditions or use-after-free vulnerabilities
would require additional reasoning about the life cycle of resources
and objects, but we are convinced that our approach and implementation take
a large step towards reaching this goal.

Note that the models we use for our supported vulnerability classes are not
exhaustive. For example, there will be instances of buffer overflow
vulnerabilities which are not covered by our model, and which we (as a
consequence) cannot generate. While we do not think that the models could ever
be exhaustive in practice, they can certainly be extended to cover more cases by extending
our current heuristics.

\begin{figure*}[t]
\centering
  \centering
  \includegraphics[width=1.0\textwidth]{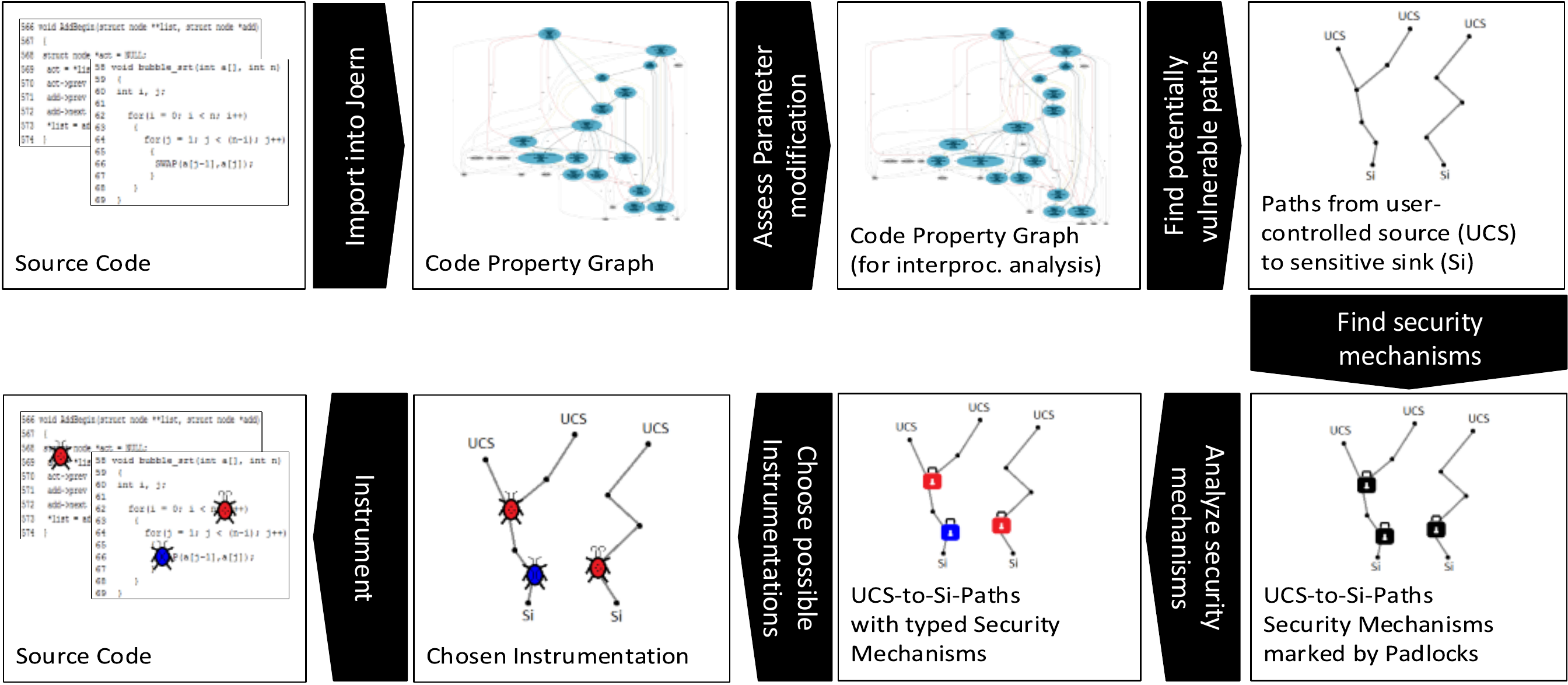}
  \caption{Workflow of Automatic Bug Insertion}
  \label{fig:workflow}
\end{figure*}

\subsection{Supported Instrumentations}

Given that taint-style vulnerabilities are defined by improperly secured data
flows and that we start with an application which we assume to be
secure\footnote{During development, we actually encountered situations where
we could not locate the security mechanism between source and sink for the
simple reason that the code was already vulnerable.}, our task is essentially
to identify and then modify the security mechanisms.

The instrumentation can fall in one of two categories:
\begin{compactenum}
\item \textbf{Invalidate security mechanisms:}
We can intentionally
weaken mechanisms, like sanitization or guard statements,
which protect a security sensitive API call.
This could mean to the protection altogether or to modify a necessary length
check to never trigger.

\item \textbf{Using security anti patterns:} We can leverage typical patterns
of vulnerable code. %
For example, there are a lot of faulty check patterns that do not
actually prevent and/or detect an integer overflow.
One could also %
transform \texttt{printf("\%s", buf)} to \texttt{printf(buf)} to introduce a
format string vulnerability.
Furthermore, TOCTTOU races~\cite{temporal-bugs} could be introduced by altering
\texttt{open()/access()} or \texttt{stat()/open()} constructs.
\end{compactenum}

We prevent syntactically incorrect code by applying our
instrumentations in a conservative manner, meaning that we do not apply it
if the specific instrumentation does not ``understand'' every element of the
security mechanism at hand.

\subsection{Workflow}

Figure~\ref{fig:workflow} shows the individual stages of our system, while Listing~\ref{lst:high-level-algo} depicts a high-level description
of our algorithm for automatic insertion of security-critical bugs:

\begin{compactenum}
\item Preparing the analysis,
\item Enabling interprocedural analysis,
\item Finding potentially vulnerable paths and finally
\item Instrumenting potentially vulnerable paths to become actually vulnerable.
\end{compactenum}
\vspace{2mm}

\begin{minipage}{0.95\columnwidth}
\lstset{language=C,
        basicstyle=\scriptsize,
        keywordstyle=\color{blue}\ttfamily,
        stringstyle=\color{red}\ttfamily,
        commentstyle=\color{green}\ttfamily,
        morecomment=[l][\color{magenta}]{\#},
        numbers = left,
        frame=lrtb,
        caption={High-Level Description of our Algorithm for Automatic Bug Insertion},
        label={lst:high-level-algo},
        captionpos=t,
        mathescape
}
\begin{lstlisting}
Find all sensitive sinks
for(each found sensitive sink):
   Trace data to user-controlled source
   Find control flows from source to sink
for(each found control flow):
   Find relevant security mechanisms in path
   if(matches known vulnerability class):
      Find applicable instrumentations
      Use randomly chosen instrumentation
\end{lstlisting}
\end{minipage}

\begin{minipage}{0.95\columnwidth}
\lstset{language=C,
        basicstyle=\scriptsize,
        keywordstyle=\color{blue}\ttfamily,
        stringstyle=\color{red}\ttfamily,
        commentstyle=\color{green}\ttfamily,
        morecomment=[l][\color{magenta}]{\#},
        numbers = left,
	showstringspaces = false,
        frame=lrtb,
        caption={Running Example for Automatic Bug Insertion},
        label={lst:running-example},
        captionpos=t,
        mathescape
}
\begin{lstlisting}
int read_from_file(FILE *f) {
 int length;
  fread((char *)&length, sizeof(int), 1, f);
  return length;
}

void wrapper(FILE *f, int *the_len) {
  *the_len = read_from_file(f);
}

void copy_buffer(  FILE *f_true, FILE *f_false
                 , char *buf, int which_file
                 , int use_wrapper) {
 int len;
  if(use_wrapper) {
   if(which_file) wrapper(f_true, &len);
   else           wrapper(f_false, &len);
  }
  else {
   if(which_file) len = read_from_file(f_true);
   else           len = read_from_file(f_false);
  }

  if(len > 256) {
   printf("ERROR: len is too big.\n");
   exit(1);
  }

 char local[256];
  memcpy(local, buf, len);
  memset(buf, 0, 512);
  do_something_with(local);
 } 
\end{lstlisting}
\end{minipage}

Throughout this section, we will use the running example depicted in
Listing~\ref{lst:running-example} to explain the individual steps.
Essentially, the shown program reads a length field from a file (line 3)
and uses it to copy up to 256 bytes of data from one buffer into another (line 30).
Furthermore, it also invalidates the original buffer (line 31)
by calling \texttt{memset()}.
For illustrative purposes, the function \texttt{copy\_buffer()} has switches
to read from two different files and optionally use a wrapper function for doing so.
While the function performing the reading returns the read value, the wrapper
function modifies its argument to show another type of data transfer.

\subsubsection{Preparing Analysis}

Starting with the source code of the application, we first invoke the preprocessor to handle
preprocessor directives, which results in pure C source code.
In the next step, we compute the \emph{code property graph}~\cite{yamaguchi:2014} from the
source code. Essentially, this means to use robust,
island-grammar-based~\cite{island-grammar} parsing to put the source code into
a graph database, which represents the abstract syntax tree (AST), types, control flow and data flow as
nodes and edges. All further analysis happens solely on this graph database
and we only turn to the source code files for later instrumentation.

\subsubsection{Enabling Interprocedural Data Flow Analysis}

The reference implementation of code property graphs does not support
interprocedural analysis out of the box.  This means for example that if a
subfunction changes one of its parameters, this is not reflected in the
data-flow graph of the calling function, where the argument to the
subfunction should be marked as modified.

\begin{figure}[t]
  \centering
  \includegraphics[width=0.15\textwidth, angle=90]{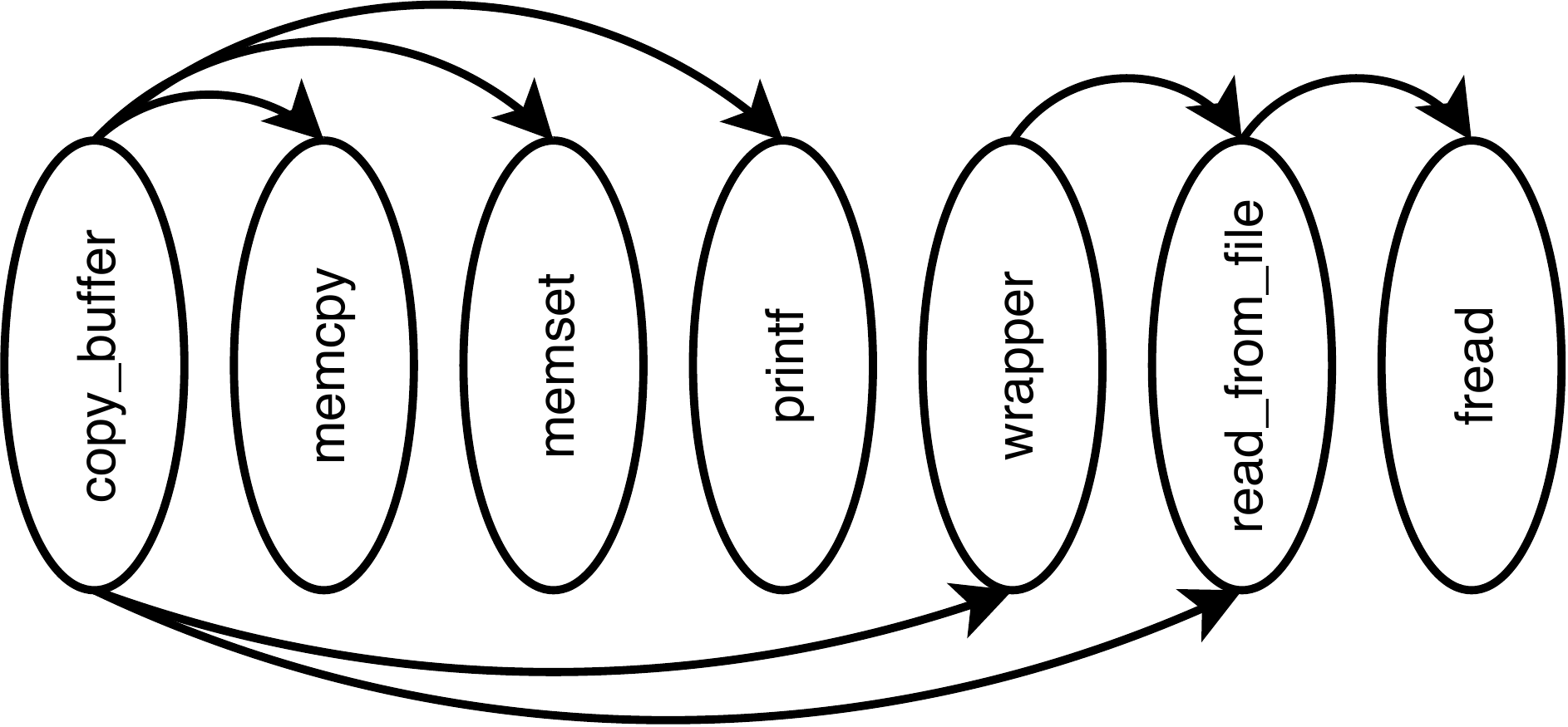}
  \vspace{5pt}
  \caption{Topological sort of the functions in Listing~\ref{lst:running-example}}
  \label{fig:topo-sort}
\end{figure}

To rectify this, we first compute the call graph of the application. %
To establish which parameters a function modifies, it is necessary to know
which subfunctions modify their parameters.  Thus, we use a topological
sorting of the graph to analyze subfunctions first. Figure~\ref{fig:topo-sort} shows
the exemplary topological sort for the running example from Listing~\ref{lst:running-example}.
In this figure, all edges are left-facing, meaning that analyzing the functions
from left to right is a valid sequence.
Once we know which parameters a subfunction modifies,
the data-flow graph in the calling function can be adjusted accordingly.
Note that static analysis does not always allow to precisely compute
if parameters are modified or not, which is why we allow a ``maybe'' state at this point.

In the running example, \texttt{copy\_buffer()} has to be augmented,
because \texttt{memset()}, \texttt{memcpy()} and \texttt{wrapper()} each modify one
of their arguments. Furthermore, every function calling \texttt{copy\_buffer()}
would have to be modified as well, because it sets the value of \texttt{buf}.

In case of circular dependencies in the call graph,
we follow a best-effort approach to break the circle by picking a function
which has the least number of calls. We expect most of
those cases to stem from simple recursive functions (i.\,e., the circle has only
two identical elements and it is irrelevant, which is analyzed first).

Naturally, we can only analyze code that is available, but especially code from external
libraries may not always be present for analysis.
Thus, we incorporate means to read their parameter modification status from
a provided file.
We handle the standard C library \texttt{glibc} in this way,
as it is linked into virtually all C programs. %

\subsubsection{Finding Potentially Vulnerable Locations}

Now that we have set up the source code for interprocedural analysis, recall
that a taint-style vulnerability is defined by an insufficiently secured data
flow between a user-controlled source and a sensitive sink.  Thus, to insert a
vulnerability, we start by finding all sensitive sinks in the application. For
us, a sensitive sink is a certain parameter of a given security-critical
function, like the length field of a \texttt{memcpy()} in line 30 of Listing~\ref{lst:running-example}.
Naturally, this
depends on the type of vulnerability: for buffer overflows, this could be
\texttt{memcpy()}, while for information leaks it could be \texttt{printf()}.
To some extent, also functions like \texttt{malloc()} are sensitive, as a
user-controlled source deciding on the number of bytes to allocate could easily
lead to a denial of service attack or maybe even cause more harm.

Next, we try to trace the data sources of a sensitive sink to a user-controlled source.
We currently define files, network connections, command-line arguments,
standard input streams (\texttt{stdin}) and environment variables to be user-controlled.
Contrary to the sinks, the user-controlled sources are not specific to the
vulnerability, but to the application.
Our choice
should cover most user-land use cases, but especially for kernels, where every
piece of data from user-land has to be regarded as potentially malicious, other
functions have to be added, such as \texttt{copy\_from\_user()} or \texttt{get\_user()}.
For now, we do not account for database interfaces,
as the types of interfaces are simply too diverse in practice.
Note that in contrast to bug finding, it is not necessary for bug insertion
to cover all possibilities,
because a missed path does not result in a vulnerability being undetected,
but only in one less opportunity for an inserted bug.

In the running example, possible data sources for the \texttt{length} value
in the \texttt{memcpy()} are in lines 16/17 (set by \texttt{wrapper()}),
as well as in lines 20/21 (set by \texttt{read\_from\_file()}).
In the third and fourth case, the value is set as the return value (line 4) of
\texttt{read\_from\_file()}. This return value carries the name \texttt{length} at this point.
In turn, the value \texttt{length} is set in line 3 as the first argument of \texttt{fread()},
which is a user-controlled source. Retracing the data source for each step,
we have now found a data flow between a user-controlled source and a sensitive sink.
For the first and second case, we have to trace the value \texttt{len} to the parameter
of \texttt{wrapper()}, where it is renamed to \texttt{the\_len} (line 7).
Apart from the different variable names, the same transitions as in third and fourth case
happen from this point on.

Once we found a data connection between a user-controlled source and a
sensitive sink, we enumerate all the control flows spanned by this data flow.
This is an important distinction, since the data flow does not include
statements which do not modify the variables in question,
but are nevertheless executed on the path.
Next, we have to find
the security mechanisms in each control-flow path. %
Each transition in a data flow is caused by a variable transferring data.
Thus, an edge between two data-flow nodes can be labeled with that variable.
Naturally, this label also overarches the control flow between the two data-flow nodes.
To find security mechanisms, we now follow the uses of the overarching variable
in the specific control-flow range. Essentially, when these variables occur
in the constraint of an \texttt{if} statement, we assume it to be a guard statement.
Sanitizations on the other hand depend on the respective types of source and sink,
which is why specialized methods would be necessary to recognize them~\cite{placing-sanitizers}.

Regarding the running example, consider the data flow for reading the first
file without using a wrapper function (line 20).
The control flow between the \texttt{memcpy()} and the call to \texttt{read\_from\_file}
traverses line 29 and (potentially) the lines 27-24 as well---even though
in practice, only line 24 will be traversed, since the execution would be aborted
in line 26. Next, line 20 is traversed, which includes a call.
However, since the value stems from a return statement, one has to retrace
the function from the end, starting at line 4 and ending in line 3.
Thus, the control flow for this source-sink pair consists of the lines
$(3,4, 20, 24, 29)$. Of those, only line $24$ holds a check and since
it is overarched by the variable in question (\texttt{len}), it is considered
a security mechanism.

\subsubsection{Instrumenting}

Once the potentially vulnerable control flow and supported security mechanisms
are found, we have to transform the underlying source code such that it is
actually vulnerable.
We start with a more careful analysis of the mechanism at hand and its surroundings,
to enumerate the number of possible instrumentations for disabling this
specific security mechanism.
For example, for a guard statement, we have to find out whether is supposed to be triggered or not.
At this point, we use heuristics depending on the presence of \texttt{return} or \texttt{exit} statements,
setting of warning or error values, or signals and exceptions.
This way, we can conservatively
remove security checks,
which would abort the execution anyways.
Thus the program
should run just as before on benign inputs.
However, the program will not reject a malformed input anymore,
but propagate it to the sensitive sink.

Once the set of applicable instrumentations is established, our prototype picks
one at random. Using the source code location information from the graph database,
we can now apply the source code transformation on the source code.

In our running example, we could modify line 24 to read
\texttt{if(wrapper == 0xDEADC0DE \&\& len > 256) \{},
which would mean that the constraint never evaluates to true,
which in turn allows values larger than 256 to be passed to the
\texttt{memcpy()}. This would result in a stack buffer overflow,
which in turn means that we have inserted a bug that
is most likely exploitable.

The high-level description of our algorithm does not account for two optimizations:
First, once it is known whether a traversed subpath for a specific variable
ends in a user-controlled source or not, one can cache this result
to prevent traversing this subpath again.
Second, to insert another bug,
one can reuse a found user-controlled source, the sensitive sink
and data-flow paths between them, while choosing another control-flow path
and another instrumentation.

One important metric for our tool would be the number of potentially vulnerable
paths it can find. However, our running example in Listing~\ref{lst:running-example}
was chosen to show that the number of such paths can be quite misleading.
In the example,
we connect the same sink (\texttt{memcpy(), line 30})
to the same source (\texttt{fread()}, line 3) in two different ways:
once through \texttt{wrapper()} and once through \texttt{read\_from\_file()} directly.
However, since we can do so in two different places, one would have to
count this as four different data flows, even though only two of them perform
different steps.
One can see that each intermediary
node in a data flow could potentiate the number of possible paths.
The same is true for the control flow in between as well.
Thus, while we report the number of paths our tool can find in the experiment
in Section~\ref{evaluation:bugdooring}, we consider the number of unique
connected source-sink pairs to be less meaningful.
Furthermore, the number of bugs we can insert depends on the number of potentially
vulnerable paths. But even if we could count them,
the number of potential instrumentations cannot be estimated in a meaningful way.
In the example from above, any expression, which evaluates to \texttt{false}
could be used. Given that there is a sheer infinite number of syntactic ways to
do that using different variables, magic constants or arithmetic operations,
we refrain from giving a number at this point.

%% file: sections/implementation.tex
\section{Implementation}
\label{implementation}

In this section, we further deepen the concepts introduced in the previous
section by explaining implementation-specific details of our prototype 
\tool{}, which we implemented using Java on a machine
running Debian 8 ``Jessie''.

\subsection{JOERN and Graph Database}

\textsc{Joern}~\cite{yamaguchi:2014} is the central component of our prototype, thus we
begin by explaining its role in our workflow.  It uses island
grammars~\cite{island-grammar} to parse C code (and to some extent C++ code) in
a robust manner, meaning that it can for example handle missing headers and
non-compiling code.  Naturally, further analysis is hindered in such situations,
but the framework will usually succeed to create meaningful output for such
partially defined code.  The resulting code property graphs, which encode the AST, control-flow
and data-flow information in annotated nodes and edges, is written into
the graph database \textsc{Neo4J}~\cite{neo4j}.

\subsection{Functions Changing Their Parameters}
\label{impl:func-sets-param}

As already mentioned, \textsc{Joern} does not take interprocedural data flows
into account. We rectify this in two steps: first, for each function, we analyze
which of its parameters it sets. Then, we augment a
function's data flow to take into account which of its subfunctions modify their arguments.
We modified \texttt{ArgumentTainter}, a tool
shipped with \textsc{Joern}, to allow batch processing functions.
In the terminology of the tool, a tainted argument
is one which gets modified. %

As mentioned above, subfunctions have to be analyzed before the function calling them.
This requires us to find
function pointers and estimate their possible values. While the former
can be implemented efficiently, determining their possible values
at each callsite is a hard task in itself. To be on the safe side, we
consider all functions, which could be assigned
anywhere in the program, to be possible values at that callsite. In essence, we
use the same algorithms to trace the data sources of function pointers that we use
for the remaining analysis. However, since a complete
data-flow graph is not available yet, we walk along the control-flow graph and
check for uses of function names and, naturally, we stop at
function-uses instead of user-controlled sources.

Note that the database still does not have
data-flow edges from one function to another. To cross function boundaries for
interprocedural analysis, the algorithm we discuss in
Section~\ref{impl:finding-paths} is necessary.

\subsection{Preprocessing}

Since we focus on vulnerabilities in C code, we do not want our analysis
to be hindered by preprocessor directives, which could disrupt the C language
semantics, e.\,g., with conditional compilation.

Thus, we execute the preprocessor on the source code files before parsing them
with \textsc{Joern}.
Given that there is a compiler flag to do this, this seems like a trivial step.
One might overlook, however, that this requires building the application, which
in turn may require a non-trivial build configuration.
To solve this problem, we take a pragmatic approach:
we build the program using the provided configuration,
using the \texttt{-{}-dry-run} option of \texttt{make},
and record the used compiler invocations. Then, we
modify the would-be executed command lines to invoke the preprocessor instead
of actually compiling or linking.

\subsection{Finding Potentially Vulnerable Source \\ Code Locations}
\label{impl:finding-paths}

We start by finding all user-controlled sources and sensitive sinks.
Given that
we opted to search our way backwards from a sink to a
source, it would be sufficient to find just the sinks to start our
search.
However, having the sources predetermined as well allows us to decide quickly
whether we reached user-controlled data or not in later phases---simply because we already made
that decision upfront.

The algorithm we use to trace data-flow information is depicted in
Listing~\ref{lst:data-trace}.
In essence, this algorithm gets a sensitive sink as input and puts it into a queue.
Then, for each node in the queue, it computes the source for that specific node
and saves it in a tree-like structure. Once it finds a node to be user-controlled,
it can traverse this definition tree back to the root to construct a data-flow path
from a user-controlled source to a sensitive sink.
Of course, the actual algorithm has to be more careful
to handle all the edge-cases and to provide fallbacks, should \textsc{Joern}
have malformed output.  Furthermore, it tracks the variable names of every traversed
data source.  Obviously, the variable names can change when one variable is
assigned to another.  However, they also change in the context of
interprocedural analysis.
This happens when a variable is passed as an argument to a subfunction.
In this case, it assumes the name of the parameter in that subfunction.
This is true in reverse, too, so that when tracking the data sources of a
parameter, the variable assumes each name of the matching argument at each
callsite.
However, it also happens, when a variable is returned, in which case the name
of the returned variable becomes the name of the assigned variable and vice versa.

Contrary to the algorithm in Listing~\ref{lst:data-trace}, our implementation
allows to find more than one user-controlled source.
It performs a breadth-first search for the data sources
of a certain node, using a queue to save nodes reached
through data flows from the start node. Then, it subsequently pops the first
node from the queue and adds nodes to the end of the queue, which are
reached by the popped node.

However, the possible complexity of C expressions makes deciding which variables
influences the current value, either by modification or by full assignment, non-trivial.
This decision is even more complex for interprocedural analysis.
Essentially, we distinguish five cases, which correspond to the steps
1.3.1 to 1.3.5 in the algorithm:

\begin{enumerate}
\itemsep 0mm
\item \textbf{Increment/Decrement:}
\item[] E.\,g., \texttt{++a;} or \texttt{(my\_struct.member\_1)-{}-;}
\item \textbf{Left-hand-side of arithmetic expression:}
\item[] E.\,g., \texttt{a = b + (c * 2);} $\Rightarrow$ \texttt{b}, \texttt{c}
\item \textbf{Assigned as return-value of function-call:}
\item[] E.\,g., given \texttt{c = f(a, b);} and
\texttt{int f(int x, int y) \{int r = x+y; return r;\}},
one has to continue at \texttt{r}, when looking for the data sources of \texttt{c}.
\item \textbf{Assigned as argument in function-call:}
\item[] E.\,g., for \texttt{strcpy(dst, src);}, the data source for
\texttt{dst} is \texttt{src}.
For functions given in the source code, we can trace the data flow from
assignments to the parameter in question, which will then end up in another parameter.
For external functions, however, we use a precomputed ``data-transfer'' lookup-table.
\item \textbf{Assigned as parameter of a function:}
\item[] When tracing the data source for the local variable \texttt{a} in
\texttt{void f(int x) \{a = x;\}}, the data flow for \texttt{a} ends at the parameter \texttt{x}.
Thus, we find all callsites of \texttt{f}, e.\,g., \texttt{f(y)}, and
continue retracing the data flow at \texttt{y}.
\end{enumerate}

While we try our best to detect assignment-by-alias,
we can obviously only do so to a limited extent,
as pointer aliasing is known to be a hard problem~\cite{pointer-aliasing-is-hard}.
Certainly, there exist modern approaches,
which are both reasonably accurate and fast~\cite{modern-points-to-1,modern-points-to-2},
but we deem our best-effort approach to be sufficient to introduce and demonstrate
the concept.

\begin{figure*}
\lstset{language=C,
        linewidth=1.0\textwidth,
        basicstyle=\scriptsize,
        keywordstyle=\color{blue}\ttfamily,
        stringstyle=\color{red}\ttfamily,
        commentstyle=\color{green}\ttfamily,
        morecomment=[l][\color{magenta}]{\#},
        numbers = none,
        frame=lrtb,
        caption={High-Level Algorithm for Tracing Data-Sources},
        label={lst:data-trace},
        captionpos=t,
        mathescape
}
\begin{lstlisting}
Input: sensitive sink
Output: Data-flow path from sensitive sink to user-controlled source

0. queue = sensitive sink
1. while(!queue.is_empty()):
   1.1 node, var = queue.pop()
   1.2 if(node.is_user_controlled()):
       1.2.1 return backtrace_definition_tree_of_node_till_sink(def_tree, node)
   1.3 switch(assigned_as):
       1.3.1 case Increment/Decrement:
               data_sources = self
       1.3.2 case Left-hand-side of arithmetic expression:
               data_sources = data sources of right-hand-side
       1.3.3 case Assigned as return-value of function-call:
               data_sources = return statements of called function
       1.3.4 case Assigned as argument in function-call:
               data_sources = data sources of assignments to parameter of called function
       1.3.5 case Assigned as parameter of a function:
               data_sources = data sources of argument at callsites of function
   1.4 definition_tree[node] += data_sources
\end{lstlisting}
\end{figure*}

As mention in Section~\ref{approach}, we consider files, network,
com\-mand-line arguments, the standard input stream, and environment variables to be
user-controlled. Therefore, we mark the respective arguments of the
\texttt{libc} functions responsible for these tasks as user-controlled sources.

Concerning the sensitive sinks, we focussed on spatial memory errors as
performed by the usual suspects like, e.\,g. \texttt{memcpy()},
\texttt{strncpy()} or \texttt{snprintf()}.  However, for information leaks, we
could also consider functions which transfer data out of the application (e.g.,
\texttt{fwrite()}, \texttt{printf()}, or \texttt{send()}) as sensitive.
Similarly,
functions like \texttt{malloc()} are
considered sensitive, as controlling their input could lead to
denial of service and more severe
attacks.

\subsection{Finding Control Flows}

Now that we have found a data flow between a user-controlled source and a
sensitive sink, we want to find the control flow connecting the data flow nodes.
Recall that we may have crossed function borders in the course of our analysis.
Since a
function can only be entered at its entry point and left through an exit
point\footnote{Technically, with constructs like \texttt{setjmp/longjmp} or
inline assembly, one can violate this property. However, they basically annul
the high-level semantics of the language.}, like a \texttt{return} statement or
the function end, our search for the control flow has to take care to respect
this property of functions.  To this end, our search has to be aware of the functions
holding the current pair of data-flow nodes, and which function calls the other.

\begin{figure}[t]
\centering
  \begin{center}
    \includegraphics[width=0.47\textwidth]{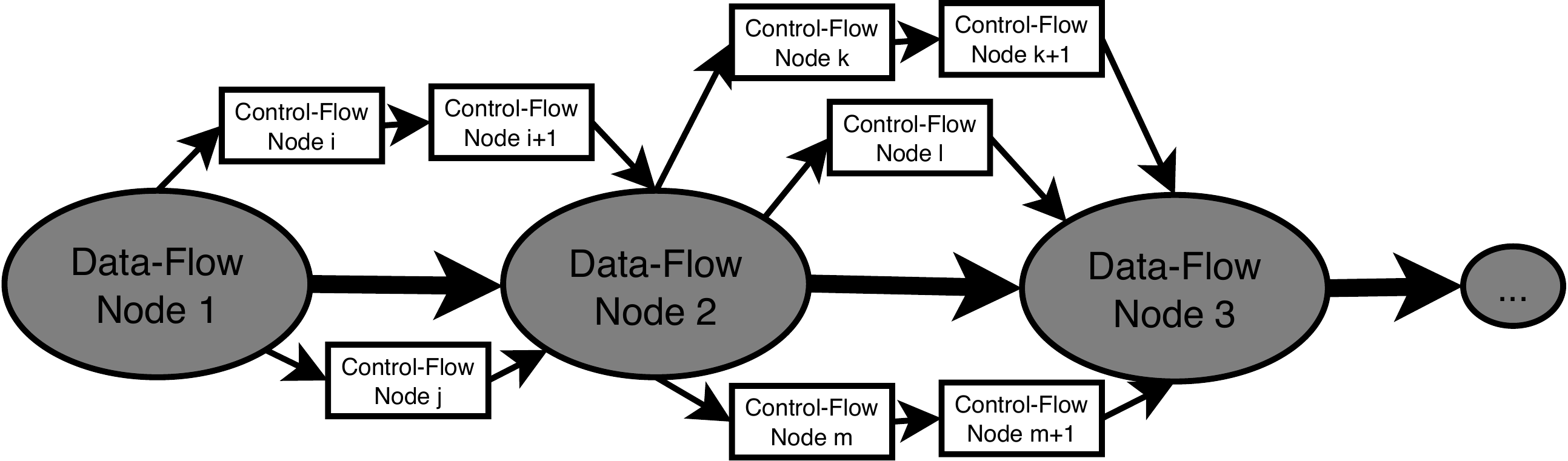}
  \end{center}
  \vspace{-5pt}
  \caption{Control flow between data-flow nodes}
  \label{fig:cfg-sub-paths}
\end{figure}

In practice, we actually want to find all control flows between two
nodes in the found data flow, instead of just one.
However, this can be done node-wise for each node in the data-flow graph,
to a data structure like the one in Figure~\ref{fig:cfg-sub-paths}.

Note that the control flow between the user-controlled source and the sensitive
sink may not stretch to the program's entry point, i.\,e., it may not
cover all the instructions, which have to be executed to later exploit the
application.

\subsection{Finding Security Mechanisms}

Having collected the possible control flows between the
user-controlled source and the sensitive sink, we can find the security
mechanisms the application uses to secure this data flow.  Since the data is
defined by the user-controlled source, previous code cannot help to do so,
just as no code after the sensitive sink can, because the
potentially harmful instruction will already have been executed.  Note that
this assumption may not hold in face of parallel execution or precautions taken
by the operating system.

We divide security mechanisms into security checks and sanitizations.
For our purposes, a security check has the following properties:
  \begin{compactenum}
    \item It occurs in the control flow between user-controlled source and sensitive sink.
    \item It uses data derived from the sensitive data ending up in the sensitive sink.
    \item Its result influences the control flow to stop this data flow to the sensitive sink.
  \end{compactenum}

\vspace{2mm}
This captures both direct and indirect examples:
\lstset{language=C,
        linewidth=1.0\textwidth,
        basicstyle=\scriptsize,
        keywordstyle=\color{blue}\ttfamily,
        stringstyle=\color{red}\ttfamily,
        commentstyle=\color{green}\ttfamily,
        morecomment=[l][\color{magenta}]{\#},
        numbers = none,
        frame=none,
        mathescape
}
\begin{lstlisting}
     fread((void *)&len, 4, 1, file);
       if(len < 256) memcpy(dst, src, len);

     bool is_too_large = strlen(argv[1]) >= 256;
       if(is_too_large) exit(1);
\end{lstlisting}

To find security checks, we simply follow the path between source and sink,
while tracing which data is influenced by the sensitive data. Then, we use
this information to determine if it causes a control-flow divergence.

For sanitization, the constructs to search for are highly specific for the
sensitive sink: data which ends in a shell-command (e.g., \texttt{system()} has
to be sanitized in a different way than data for an SQL query.  One cannot,
however, assume in general that every change to the data is actually a
sanitization.
As of now, our prototype does recognize some sanitizations, like inserting
\texttt{0x00}-bytes into strings to restrict their length,
but our instrumentations do not cover manipulating such sanitizations, yet.

\subsection{Instrumentations}

The instrumentations, i.e., our code transformations that actually disable security mechanisms,
happen in three steps:

\begin{compactenum}
   \item \textbf{Bugdoorability:} Decide, if our models ``understand'' the security mechanism.
   \item \textbf{Applicable instrumentations:} Enumerate the possible instrumentations for this security mechanism.
   \item \textbf{Apply random instrumentation:} Choose a possible instrumentation at random and apply it.
\end{compactenum}

\vspace{2mm}
The first and second steps are somewhat intermingled and rather straightforward,
as we simply compare the type of mechanisms a specific instrumentation can disable
against the security mechanism. For our prototype, we mainly focus on disabling security checks, which means
constructs like \texttt{if(length > 256) \{$\ldots$\}} and thus
implemented the following instrumentations classes:

\begin{compactitem}
  \item Remove security mechanism
  \item Surround by \texttt{if} with constraints always evaluating to false (resp. true), to never (resp. always) execute
  \item Arithmetically influence decision logic
  \item Move security mechanism into an unrelated path
  \item Swap the security mechanism and sensitive sink
  \item Use security anti-patterns for integer overflow checks
\end{compactitem}
\vspace{2mm}

Note that an instrumentation class can cover many instrumentations.
For example, we can transform a statement such as \texttt{(length > 256)} to different representations:

\begin{compactitem}
    \item \texttt{(length > 512)}
    \item \texttt{(length/2 > 256)}
    \item \texttt{(length > 256*2)}
    \item \texttt{((char)length > 256)}
\end{compactitem}

\vspace{2mm}
Thus, we do not think that there is a fair way to count the
number of actual instrumentations to introduce a bug, as instrumentation
subclasses could be defined almost arbitrarily and syntactic
changes introduced by an instrumentation class are virtually unlimited.
Thus, we refrain from reporting the number of possible
instrumentations during the evaluation of our prototype.

%% file: sections/evaluation.tex
\section{Evaluation}
\label{evaluation}

In the following, we evaluate our prototype for automatic bug insertion called
\tool{} with respect to its performance, the quantity of security-critical bugs
it can introduce, and the exploitability of said introduced bugs.

To this end, we tried to introduce vulnerabilities into open-source projects,
namely \texttt{libpng}, \texttt{wget}, \texttt{busybox}, and \texttt{vsftpd}.
We found between 22 and 158 unique source-sink pairs for each tested project,
which translates to hundreds of security-critical data paths. This in turn
implies hundreds of different security-critical bug variations we could introduce.

Our rationale regarding the exploitability is as follows:
By definition, the vulnerabilities we study have a
data flow from a user-controlled source to a sensitive sink.
If only insufficient security mechanisms are found on this data-flow path,
we have a taint-style vulnerability.
Assuming that the path was not vulnerable to begin with,
the security mechanisms on the path are the reason as to why the path is not vulnerable.
Consequently, this means that relaxing the security mechanisms makes the path
vulnerable.

Naturally, the sheer volume of generated bugs, combined with the lack
of practical automatic exploit generation, means that we cannot verify
that every single introduced bug is actually exploitable.
However, we chose to justify our claim of introducing exploitable bugs by taking
a patched application, which had real-world exploitable CVEs,
and re-introducing a security-critical bug into the path in question.

\subsection{Setup}

All experiments were conducted in a virtual Debian 8 ``Jessie'' machine,
with an
Intel Core i7-2640M @ 2.8GHz, 8GB DDR3-RAM @ 1600MHz and an SSD.

\subsection{Bugdooring Open-Source Projects}
\label{evaluation:bugdooring}

For this experiment, we tried to insert bugs into four open-source projects.
We chose \texttt{libpng}, the official reference library for the popular image format PNG,
because it suffered from a well-studied CVE which we wanted to use as a case study in a following experiment (see Section~\ref{eval:libpng-use-case}).
Furthermore, we chose the two popular standalone programs \texttt{wget v1.16}
and \texttt{busybox v1.24.1}. %
The FTP server \texttt{vsftpd}
was implemented with security in mind,
which is why it performs all buffer-related operations through a set of
wrapper functions to shield the actual application code from the sanity checks
necessary to prevent buffer overflows or overreads.

\begin{table}
\centering
\caption{Results of automatic bug insertion} %
\label{tab:bugdooring}
\begin{scriptsize}
\setlength{\tabcolsep}{2.5pt}
\begin{tabular}{l|r|r|r|r}
 & \textbf{libpng} & \textbf{vsftpd} & \textbf{wget} & \textbf{busybox} \\
\hline
Lines of code & 40,044 & 20,046 & 137,234 & 265,887\\
User-controlled sources (UCS) & 9 & 3 & 21 & 152\\
Sensitive sinks (Si) & 98 & 13 &  453 & 573\\
Unique UCS-Si combinations & 158 & 22 & 22 & 30 \\
UCS-to-Si data-flow paths & 22,516 & 786 & 1,882 & 2,905 \\
\end{tabular}
\end{scriptsize}
\end{table}

As one can see in Table~\ref{tab:bugdooring}, our tool is able to find a
substantial number of sensitive data paths for each application.
The library \texttt{libpng} has a lot of such paths compared to its code size, 
which was expected since the PNG file format incorporates several context sensitive
length fields.
For this library, the number of user-controlled sources is rather low,
since most data flow actually traces back to one of a few file reading operations.
Compared to that,
the number of user-controlled sources in \texttt{wget} and \texttt{busybox}
is rather high, given that they exchange data not only via files.
Considering that we mostly searched for user-controlled length fields,
the number of found potentially vulnerable paths was expected.
As mentioned before, \texttt{vsftpd} performs all buffer-related operations
through a set of wrapper functions.
Indeed, the number of both user-controlled sources and sensitive sinks
is very low, and all the potentially vulnerable paths our current implementation found
traverse the security checks of these wrapper functions.

The value for unique user-controlled source (UCS) and sensitive sink (Si) combinations
refers to the number of user-controlled source for which we found data-flow
from a single sensitive sink, summed over all sensitive sinks:
\[
\sum_{s \in Si}{\#\{u \in UCS, \text{ where } s \text{ connects to } u}\}
\]
In contrast, the number of UCS-to-Si data-flow paths refers to the sum over all
data-flow paths that connect the found user-controlled sources and sensitive sinks.
Since \textsc{Joern} sometimes includes transitive data-flow edges,
the reported
numbers may be higher than the data flows occurring when executing the program.
Thus, they can be seen as an upper bound, while the unique
UCS-Si combinations can serve as a lower bound, which means that the actual number
of data flows can be estimated to be between those two.
Since the number of control-flow nodes connecting to data-flow nodes suffers
from the same problem, we do not report them here.

As mentioned before, there is no fair way to estimate the number of applicable
instrumentations for a given UCS-to-Si path, but given that %
multiple instrumentation classes are applicable to each found path,
it is safe to say that we can introduce at least a hundred bugs into each
tested project.

\subsection{Performance}

\begin{table}
\centering
\caption{Runtime in minutes:seconds for different phases of processing open-source projects}
\label{tab:runtime}
\begin{scriptsize}
\begin{tabular}{lrrrr}
\textbf{Runtime for phase} & \textbf{libpng} & \textbf{vsftpd} & \textbf{wget} & \textbf{busybox} \\
\hline
\multicolumn{5}{l}{Importing with \textsc{Joern}} \\
 & 00:32 & 00:24 & 00:58 & 03:49\\
\\
\multicolumn{5}{l}{Analyzing intraprocedural behaviour} \\
 & 00:35 & 00:12 & 01:33 & 03:42\\
\\
\multicolumn{5}{l}{Augmenting code property graph for interprocedural analysis} \\
 & 01:18 & 00:22 & 03:40 & 35:28 \\
\\
\multicolumn{5}{l}{Finding UCS-to-Si paths} \\
 & 01:56 & 00:49 & 02:14 & 7:05 \\
\hline
\textbf{Total:} & 04:21 & 01:47 & 08:25 & 50:04 \\
\textbf{...per 10,000 LOC:} & 01:05 & 00:53 & 00:37 & 01:56 \\
\end{tabular}
\end{scriptsize}
\end{table}

Table~\ref{tab:runtime} shows that
transforming the source code into code property graphs using \textsc{Joern} is fast,
although it does not behave linearly in the size of code base.
The next step (i.e., analyzing which functions set which parameters) is slower,
but seems to increase linearly with the size of the code base.
Naturally, some functions take longer to analyze than others,
depending on the number of their parameters, the number of subfunctions they invoke
and their general size, but we observed this to average out over the full code base.
The time for augmenting the code property graph to facilitate interprocedural analysis,
as expected, grows for larger code bases,
but the size of the code base is not the only relevant factor:
we observed that intertwined programs, where a lot of function are
called in a lot of places, result in increased runtime.
Similarly, the time for finding potentially vulnerable paths depends not
only on the size of the code base, but especially on the fan-out of potential
continuations of the data flow.
Function pointers in particular are problematic in this context.
Since a specific data-flow path only has to be traversed once per
variable to know whether it ends up in a user-controlled source, %
tracing the data flow for a specific sensitive sink tends to get faster
towards the end of the algorithm when more nodes have already been visited.

Given the nature of C code, transforming it into property graphs is most likely
best performed in one step. Intraprocedural analysis to facilitate the
augmentation of the code property graphs for interprocedural analysis, however,
could be performed in an ad-hoc fashion, depending on the necessities of the
source-to-sink data-flow analysis.
This would allow us to analyze a smaller subset of the program and increase performance.

\subsection{Case Study: Libpng CVE-2004-0597}
\label{eval:re-introduce-cves}
\label{eval:libpng-use-case}

Libpng %
had multiple buffer overflow vulnerabilities described in CVE-2004-0597.
Here, we will discuss the automatic removal of the introduced security guards.
We chose this example for two reasons:
First, it is well studied, as its exploitation is a training example in an undergrad course~\cite{bratus-abacus}.
Second, it highlights some of our tool's features, namely
interprocedural analysis,
function pointers,
data-flow through struct members,
and removal of multiple guards.

In particular this CVE describes a stack buffer overflow due to a malformed palette-index. %
A stack buffer of fixed size (256 bytes) is allocated to hold a color palette index.
However, the length of the palette is read from a png file and libpng only checked
whether this value exceeded the size of the main color palette.
This did not prohibit the attacker from claiming an index palette size of
more than 256 bytes in the png file, which lead to a stack buffer overflow.
Thus, a patch was issued to additionally check, whether the length exceeded
256 bytes, the maximum size for an index palette.

\subsubsection{Sensitive Sink and User-Controlled Source}
The length for the palette-index is passed through four functions in three
different files, until the buffer is filled at the sensitive sink: a call to \texttt{fread()}.
Since \texttt{fread()} is user-controlled,
the attacker controls the stackbuffer's contents,
which makes this vulnerability likely to be exploitable.

The way to the user-controlled source for the length-field is
a little more complex.
Over the course of ten functions in five files, it changed its name eight times
and its type once.
One had to track data flow through three different struct members,
the data was copied, passed as parameters, returned
and filled via \texttt{memcpy()} from an internal buffer,
which in turn was finally set via \texttt{fread()}. %

\subsubsection{Instrumentations}

There are 22 nodes in the data-flow path from the user-controlled source to the sensitive sink,
but the shortest of the 144 connecting control-flows is already 114
nodes long and includes 37 checks.
Only five of those are overarched by the respective variable connecting two
nodes in the data flow.
Three of them check internal relations between buffer sizes
and another one checks for an out-of-range integer.
The final one is the check which was issued with the patch.

The conditions of the first three checks have to be fulfilled to continue execution,
but their alternative path does not immediately abort the program
and thus our approach deems them not to be security critical.
The out-of-range check, however, would abort execution,
which is why our approach would instrument it to never evaluate to true.
While this introduces a bug,
it does not hinder execution on well-formed input.
As for the last check, our tool automatically determines that it
must never evaluate to true and can determine which of our instrumentation
classes apply. Then, it applies one of those instrumentations
and thereby introduces a bug. Because we know that an exploit for the program
lacking this very check exists, we can conclude that our tool
successfully inserted an exploitable vulnerability.

\paragraph{General Obstacles}

We found that \texttt{libpng} uses the C~preprocessor directive
\texttt{\#define png\_memcpy memcpy}
and only uses \texttt{png\_memcpy()} throughout its code.
We found this pattern of using macros for simple wrapper functionality to occur frequently.
Since this alias does not stem from C code, it cannot be detected with C-analysis alone.
Thus, source code analysis for C either has to be aware of preprocessor macros
or utilize the preprocessor, like we do.

%% file: sections/discussion.tex
\section{Limitations and Future Work}
\label{discussion}

We now discuss limitations of our approach and current prototype,
possible future work, and alternative use cases of automated bug insertion.

\subsection{Exploitability}

Our technique finds path between user-controlled sources and
sensitive sinks, and then modifies or removes security mechanisms on these paths,
thus creating security-relevant bugs. However, this may not be enough to
actually create an exploitable vulnerability, as we
cannot assert the
global satisfiability of all path conditions, which are necessary to traverse
the path in question.

One observation argues in favour of possible exploitation despite all this: the
security mechanisms are present in the program. If no values existed for
traversing the path in question, the security mechanism would be superfluous in
the first place.
Note that this assumes that no overly defensive programming strategy was used.

While research towards automatic generation of exploits~\cite{aeg,mayhem} or
at least some proof of seriousness of the bug~\cite{convicting-exploits}
exists, verifying all generated bugs with such complicated additional
components seemed unreasonable.  As mentioned in
Section~\ref{eval:re-introduce-cves}, we tried to justify our claim of
exploitability by letting our tool reintroduce known-to-be-exploitable bugs.
Nevertheless, automatic satisfiability verification and exploit generation are
out of scope for this paper.

Unfortunately, the lack of confirmed vulnerabilities in combination with the
impossibility to count the number of introduced bugs in a meaningful way means
that it is not possible to report something like a false positive rate for our
approach.

\subsection{Additional Vulnerabilities}

The bugs we introduce are limited in two ways: first, we only support a
limited number of vulnerability types, which all belong to the class of
taint-style vulnerabilities, as they allow rather conventional exploitation.
Thus, we cannot introduce other types of bugs for now.  However,
we believe implementing additional bug classes to be straightforward.

Second, while we do add some element of randomness to the introduction of bugs,
they undoubtedly have a pattern. For the use-case of an artificial bug corpus,
this might arguably be problematic, as it would be a very valid strategy to
model the heuristics we used to introduce the bugs to find them later on.
However, given that we can automatically introduce such bugs, it can also be
argued that finding these bugs, whether they have a pattern or not, is
absolutely mandatory. Hence, we create some kind of baseline to
evaluate techniques that aim to detect vulnerabilities in an automated manner.

Furthermore, inserting additional vulnerable paths or functions, instead
of only weakening present security mechanisms,
would also be interesting for future work.

\subsection{Alternative Use Cases}
\label{discussion:usecases}

We focused on the generation of test corpora in this paper, but we do see other
use cases for bug insertion.
Capture-the-flag (CTF) contests essentially pose exploiting challenges,
for which vulnerable programs are a necessity. While our approach does not
guarantee exploitability and is not (yet) targeted towards vulnerabilities
which are tricky to exploit, we think that it could be a valuable tool
for the organizers.

Furthermore, the ability to insert exploitable bugs could theoretically be used
to facilitate later exploitation.
However, in this scenario, the attacker would require write access to the
source code. While a recent publication~\cite{omitting-commits} shows that
tampering with version control systems is feasible, it is a big obstacle.
Furthermore, the inserted vulnerability should be hard to find and exploitable
for a long time, i.\,e., not be removed soon.
Given that we want to insert many vulnerabilities instead of a single,
special one, we think that manual effort would be the way to go for an attacker,
and thus do not see an ethical problem with our approach.

%% file: sections/related_work.tex
\section{Related Work}
\label{related-work}

The work most closest to our approach is a recently published paper entitled
\textit{LAVA: Large-scale Automated Vulnerability Addition}~\cite{lava}.
The authors want to generate a sufficient number of bugs for purposes of
testing bug-finding tools.
In contrast to our approach, they chose a dynamic method by tainting
input bytes and tracing them through the program. They look specifically
for rarely modified and dead data, for which they then insert code
performing either a buffer overread or buffer overflow. If necessary, they %
introduce new static or global variables to allow the needed data flow.
Additionally, they insert guards to execute the vulnerability only if a magic
value occurs in the input.
As the name suggests, one could say that they actually \textit{add} new
vulnerabilities, while we transform code from invulnerable to vulnerable,
i.\,e., insert bugs.
We have chosen a different
methodology, so that we deem this to be concurrent and independent work.

Given the dynamic approach, LAVA also generates inputs triggering
the vulnerabilities and the authors also provide preliminary results showing that
state-of-the-art fuzzers and symbolic execution engines are not able to find
all the bugs they are able to add.
We think that this finding underlines the importance of automated bug insertion.

\subsection{Insufficient Test Data}
Miller~\cite{fuzz-by-number} uses a set of 16 hand-written vulnerabilities
to compare eight fuzzers and states that, while the scarcity of test cases is
a problem, these 16 artificial test cases already offered a lot of insight.

Nilson et al.~\cite{bugbox} state that
``existing sources of vulnerability data did not supply the necessary structure
or metadata to evaluate them completely'',
which is why they developed \textsc{BugBox}, a simulation environment with an
accompanying corpus of vulnerabilities and exploits.
Their vulnerabilities are real-world examples specific to PHP,
and they focus mostly on the aspects of exploiting.

Delaitre et al.~\cite{evaluating-bug-finders}
evaluated 14 static analyzers. %
They establish three critical characteristics for vulnerability test cases and
state that ``Test cases with all three attributes are out of reach'':
\begin{enumerate}
  \item \textbf{Statistical significance:} There must be many, diverse vulnerabilities.
  \item \textbf{Ground truth:} The location of the vulnerabilities must be known.
  \item \textbf{Relevance:} The vulnerabilities must be representative for those found in real source code.
\end{enumerate}
Test corpora generated with our approach fulfill the first two characteristics,
and we are certain that the third characteristic can be fulfilled
with carefully stated bug models as well,
given that the instrumented code stems from real programs.

\subsection{Vulnerability Databases}

According to Nilson et al.~\cite{bugbox} and Delaitre et al.~\cite{evaluating-bug-finders},
the existing databases are not
sufficient for a comprehensive evaluation of bug finding techniques.
However, that is not to say that there are no such databases.

First of all, the Common Vulnerabilities and Exposures (CVE)
and the Common Weakness Enumeration (CWE) are to mention.
The former consists of a short description of real-world vulnerabilities
and links to further resources regarding the specific vulnerability.
Its main purpose is to identify a certain vulnerability unambiguously,
but it does usually not include actual vulnerable code.
The latter often offers a few code snippets to illustrate the
hierarchized vulnerabilities, but those are likely
not sufficient for a comprehensive evaluation of bug finding techniques.

Specific to web development, OWASP WebGoat~\cite{webgoat} as well as
SecuriBench~\cite{securibench} collect vulnerabilities for illustrative purposes.
However, they do not offer a structured corpus, which is necessary for evaluation purposes.

The most useful public database for the evaluation of bug finding techniques
is generated by the NIST project Software Assurance Metrics And Tool Evaluation (SAMATE)~\cite{samate}.
Its largest standalone test suite actually contains over 60,000
vulnerable synthetic test cases, but was uploaded in 2013.
Naturally, these test suites are static and cannot generate fresh bugs.
The project also included the 
IARPA program Securely Taking on Software of Uncertain Provenance (STONESOUP)~\cite{stonesoup},
which provides 164 Java and C snippets, which can be inserted into other programs
to make them vulnerable. However, the snippets are static and require rather
specific environments.

\subsection{Mutation Testing}
Mutation testing~\cite{mutation-testing} randomly modifies the source code
to make it behave slightly differently at runtime.
In that way, both the ability of the test set to catch such errors as well as
the necessity and import of the modified piece of code can be estimated.
As a result, both coverage and overlap of modified code and test set are the
important metrics.
While, in principle, bugs like the ones introduced by our approach
could be inserted by mutation testing as well,
we purposefully insert special bugs at carefully selected locations to
introduce vulnerabilities.

%% file: sections/conclusions.tex
\section{Conclusions}
\label{conclusions}

In this paper, we proposed an approach for automatic generation of bug-ridden
test corpora. Our prototype implementation of this concept currently targets
the insertion of spatial memory errors by modifying security checks using six
different instrumentation classes. With such test corpora, we aim to facilitate
future research in the field of bug finding techniques, so that they can be
evaluated and compared in an objective and statistically meaningful way.

%% file: sections/acknowledgement.tex
\section*{Acknowledgment}

The project leading to this application has received funding from the European
Research Council (ERC) under the European Union's Horizon 2020 research and
innovation programme (grant agreement No. 640110 -- BASTION). This work was
also supported by the German Research Foundation (DFG) research training group
UbiCrypt (GRK 1817).  The authors would like to thank Engin Kirda, William
Robertson, Patrick Carter, Timothy Leek, Patrick Hulin, and Brendan
Dolan-Gavitt for the fruitful discussions. Furthermore, we thank Jan Teske and
Tilman Bender for supporting our implementation efforts.